\documentclass[11pt]{article}

\makeatletter%
\def\nottoobig#1{{\hbox{$\left#1\vcenter to1.111\ht\strutbox{}\right.\n@space$}}}
\makeatother%

\makeatletter%
\def\@begintheorem#1#2{\trivlist\item[\hskip\labelsep{\bf #1\ #2}]}
\makeatother
\makeatletter %

\newcommand{\implies}{\:\Rightarrow\:}

\newlength{\filength}
\newsavebox{\gcbox}
\sbox{\gcbox}{\framebox[\filength]{\rule{0ex}{2ex}}}

\newlength{\leftjustindent}
\newlength{\@leftjustindent}
\def\leftjust{\let\\\@leftjustcr\let\end\@endleftjust
  \addtolength{\@leftjustindent}{\leftjustindent}
  \vcenter\bgroup
  \halign\bgroup
    \hbox to\displaywidth{
      \rule{\@leftjustindent}{0ex}$\displaystyle##$\hfill
      }\crcr
}
\def\endleftjust{\crcr\egroup\egroup\endgroup}
\def\@endleftjust#1{\crcr\egroup\egroup\@checkend{#1}\endgroup}
\def\@leftjustcr{\crcr}

\newcommand{\red}[3]{ {  {\rm R}_{#1}^{#2}({#3}) }    }

\newtheorem{theorem}{Theorem}[section]

\newcommand{\qedblob}{\mbox{\rule[-1.5pt]{5pt}{10.5pt}}}
\def\literalqed{{\ \nolinebreak\hfill\mbox{\qedblob\quad}}}

\def\qed{\literalqed}

\newcommand{\singlespacing}{\let\CS=
\@currsize\renewcommand{\baselinestretch}{1}\tiny\CS}
\newcommand{\singlespacingplus}{\let\CS=
\@currsize\renewcommand{\baselinestretch}{1.25}\tiny\CS}
\newcommand{\doublespacing}{\let\CS=
\@currsize\renewcommand{\baselinestretch}{1.75}\tiny\CS}
\newcommand{\draftspacing}{\let\CS=
\@currsize\renewcommand{\baselinestretch}{2.0}\tiny\CS}

\makeatother%

\makeatletter
\clubpenalty=\@highpenalty
\widowpenalty=\@highpenalty
\makeatother

\makeatletter
\newcommand{\niceonespacing}{\let\CS=\@currsize\renewcommand{\baselinestretch}{1.1}\tiny\CS}\newcommand{\nicetwospacing}{\let\CS=\@currsize\renewcommand{\baselinestretch}{1.2}\tiny\CS}
\newcommand{\nicethreespacing}{\let\CS=\@currsize\renewcommand{\baselinestretch}{1.3}\tiny\CS}
\newcommand{\singlespacingplusplus}{\let\CS=\@currsize\renewcommand{\baselinestretch}{1.35}\tiny\CS}
\newcommand{\nicefivespacing}{\let\CS=\@currsize\renewcommand{\baselinestretch}{1.5}\tiny\CS}
\newcommand{\nicesixspacing}{\let\CS=\@currsize\renewcommand{\baselinestretch}{1.6}\tiny\CS}
\newcommand{\nicefoospacing}{\let\CS=\@currsize\renewcommand{\baselinestretch}{1.1}\tiny\CS}
\makeatother

\makeatletter
\def\@cite#1#2{[#1\if@tempswa , #2\fi]}
\makeatother

\makeatletter
\def\@citex[#1]#2{\if@filesw\immediate\write\@auxout{\string\citation{#2}}\fi
  \def\@citea{}\@cite{\@for\@citeb:=#2\do
    {\@citea\def\@citea{,\linebreak[0]}\@ifundefined
       {b@\@citeb}{{\bf ?}\@warning
       {Citation `\@citeb' on page \thepage \space undefined}}%
\hbox{\csname b@\@citeb\endcsname}}}{#1}}
\makeatother

\newcommand{\up}{{\rm UP}}

\newcommand{\p}{{\rm P}}

\newcommand{\np}{{\rm NP}}

\newcommand{\pp}{{\rm PP}}
\newcommand{\bpp}{{\rm BPP}}
\newcommand{\zpp}{{\rm ZPP}}

\newcommand{\conp}{{\rm coNP}}

\newcommand{\thetatwo}{{\Theta_2^p}}

\newcommand{\ph}{{\rm PH}}

\nicefoospacing

\def\pair#1{{{\langle\!\!~#1~\!\!\rangle}}}

\newcommand{\condition}{\,\nottoobig{|}\:}

\newcommand{\parallelnp}{\mbox{$\p_{||}^{\np}$}}
\newcommand{\rp}{\rm R}
\newcommand{\corp}{{\rm coR}}
\newcommand{\ceqp}{{\rm C_{\!=\!}P }}

\newcommand{\ppoly}{\rm P/poly}

\newcommand{\dw}{\mbox{\tt Carroll Winner}}

\newcommand{\mee}{\mbox{\tt MEE}}
\begin{document}

\bibliographystyle{alpha}

\title{Raising NP Lower Bounds to Parallel NP Lower 
Bounds\protect\thanks{%
\protect\singlespacing
Supported in part 
by grants
NSF-INT-9513368/\protect\linebreak[0]DAAD-315-PRO-fo-ab,
NSF-CCR-9322513,
and 
a University of Rochester Bridging Fellowship.}}

\author{%
Edith Hemaspaandra\protect\thanks{%
\protect\singlespacing
{\tt edith@bamboo.lemoyne.edu}.
Dept.~of Mathematics, 
Le Moyne College,
Syracuse, NY 13214, 
USA.} 
\and
Lane A. Hemaspaandra\protect\thanks{%
\protect\singlespacing
{\tt lane@cs.rochester.edu}.
Dept.~of Computer Science,
University of Rochester,
Rochester, NY 14627, 
USA.} 
\and
J\protect\"org Rothe\protect\thanks{%
\protect\singlespacing
{\tt rothe@informatik.uni-jena.de}.
Inst.~f\protect\"ur Informatik,
Friedrich-Schiller-Universit\protect\"at Jena,
07743 Jena, 
Germany.  Work done in part while 
visiting Le Moyne College.}}

\date{}

{\singlespacing

\maketitle}
{\singlespacing 

\begin{abstract}
A decade ago, a beautiful paper by Wagner~\cite{wag:j:more-on-bh}
developed a ``toolkit'' that in certain cases allows one to prove
problems hard for parallel access to~NP\@.  However, the problems his
toolkit applies to most directly are not overly natural. During the
past year, problems that previously were known only to be NP-hard 
or coNP-hard have
been shown to be hard even for the class of sets solvable via parallel
access to~NP\@.  Many of these problems are longstanding and extremely
natural, such as the {\tt Minimum} {\tt Equivalent} {\tt Expression}
problem~\cite{gar-joh:b:int} (which was the original motivation for
creating the polynomial hierarchy), the problem of determining the
winner in the election system introduced by Lewis Carroll
in~1876~\cite{dod:unpubMAYBE:dodgson-voting-system}, and the problem
of determining on which inputs heuristic algorithms perform well.
In the present article, we survey this recent progress in raising 
lower bounds.
\end{abstract}

}

\section{Introduction}

Suppose you are given some nice, challenging problem and you are 
able to prove an NP-hardness 
lower bound for it.\footnote{\protect\singlespacing
We use hardness in the sense in which it 
is becoming most commonly
used, namely, with regard to many-one 
polynomial-time ($\leq_m^p$) reductions.
That is, a problem $A$ is said to be {\em hard\/} 
for a complexity class ${\cal C}$ (for short, {\em ${\cal C}$-hard\/}) 
if every problem $B \in {\cal C}$
polynomial-time many-one reduces to~$A$, i.e., there is a polynomial-time
computable function $f$ such that, 
for each $x$, $x \in A \iff f(x) \in B$.
If $A$ is hard for 
${\cal C}$ and $A \in {\cal C}$, then $A$ is 
said to be {\em ${\cal C}$-complete}.}
Unfortunately, this lower bound doesn't match the
problem's known upper bound, which is, say, $\p^{\np}$---that is, your
problem can be solved by some deterministic polynomial-time Turing 
machine that is given access to an NP database. 
Now, when trying to close the gap between the upper and the
lower bounds, the first thing you notice is that to
solve your problem, you do not
need the power of sequential access to the NP database; instead of asking
your queries {\em sequentially\/} (which means that the answers to earlier
questions may determine later questions---as in, for example, the 
popular game Twenty Questions or as in binary search), it suffices to
ask all queries in {\em parallel\/} (i.e., on a given input~$x$, 
some P machine first computes a list of all questions that it wants
answered, then it gives its NP database 
the entire list at once, 
and finally, given as the reply 
a list of yes/no answers corresponding to the listed
questions, the~P machine correctly determines whether or not 
$x$ is an instance belonging to the given problem).

The type of restricted access to NP that
you have discovered that your 
problem can be solved with has been studied in the literature.
In particular,
Papadimitriou and
Zachos~\cite{pap-zac:c:two-remarks} introduced and discussed the
complexity class~$\parallelnp$, which by definition 
contains exactly those
problems that can be solved
via parallel access to~NP\@.  Clearly, $\np \subseteq \parallelnp
\subseteq \p^{\np}$.  
Hemaspaandra~\cite{hem:c:sky-stoc} 
and K\"obler, Sch\"oning, and Wagner~\cite{koe-sch-wag:j:diff}
proved that 
$\parallelnp$ is also exactly 
the class of problems that can be solved
by ${\cal O}(\log n)$ sequential Turing queries to
NP\@.  Wagner~\cite{wag:j:bounded}
established
about half a
dozen other characterizations 
of $\parallelnp$.\footnote{\protect\singlespacing 
For historical sticklers, we mention that this paragraph is
slightly ahistorical.  In fact, Papadimitriou and Zachos 
used as their definition what we above have described as 
the equivalent characterization of the class, and the 
papers mentioned actually linked their original definition
to what is here described as the definition.  Just to make things
more confusing, the class itself nowadays is most 
often referred to in
the technical literature as $\thetatwo$, rather than as 
$\parallelnp$ or $\p^{ \np [ {\cal O} ( \log n ) ]}$, as it has 
become a ``named'' level of the polynomial hierarchy
(see~\cite{wag:j:bounded}).}
Furthermore, it is known that if NP contains some
problem that is hard for $\parallelnp$, then the polynomial hierarchy
collapses to NP\@.  The class $\parallelnp$ is also closely
related to whether NP has sparse Turing-hard 
sets~\cite{kad:j:pnplog}, to whether feasible complexity classes
can create Kolmogorov-random objects~\cite{hem-wec:j:man-rand},
and to many other topics (see, e.g., 
\cite{lon-she:j:low,kre:j:optimization}).

Can you completely close the gap between your problem's upper and  
lower bounds by raising its NP-hardness lower bound to a 
$\parallelnp$-hardness lower bound? 
Wagner~\cite{wag:j:more-on-bh}
provided a very useful toolkit with which one can seek to 
establish
$\parallelnp$-hardness results.
In particular,
he showed how to obtain certain ``standardized'' $\parallelnp$-complete
versions associated with 
some given NP-complete problem such as {\tt Clique}, 
and he established a sufficient condition for 
$\parallelnp$-hardness. 
Using this approach, he 
showed, for instance, that the following problem is 
$\parallelnp$-hard (and, indeed, 
$\parallelnp$-complete): Given a 
graph $G$, is the size of $G$'s maximum-sized cliques
an odd integer?
Unfortunately, such standard versions of $\parallelnp$-complete
problems are arguably
not too natural (though in fairness we should mention
that Wagner's ``equality'' and ``comparison'' problems are potentially
more arguably natural).
During the past year, Wagner's
toolkit has been used to raise to $\parallelnp$-hardness the lower bounds 
of a number of clearly natural, longstanding problems for which
previously only NP-hardness (or coNP-hardness) lower bounds were known. 
In this article, we survey this progress.

In particular, we will focus on three problems, $\dw$, {\tt Minimum}
{\tt Equivalent} {\tt Expression}, and the problem of 
determining for which graphs the 
Minimum Degree Greedy
Algorithm
finds an independent set whose size is 
within a factor of~$r$ (with $r$ fixed) of the size of the 
maximum independent sets of the graph (formal 
definitions will be given later).
These problems come from 
three different areas: political science, 
logic, and graph theory.  We now briefly describe these problems
and sketch their historical background.

In 1876, Charles L.
Dodgson~\cite{dod:unpubMAYBE:dodgson-voting-system} (who is 
usually referred to today by his pen name,
Lewis Carroll) proposed an election system
in which the winner is the candidate who with the fewest changes in
voters' preferences becomes a Condorcet winner---a candidate who beats
all other candidates in pairwise majority-rule elections.  $\dw$ is
the problem of whether a distinguished candidate wins a given election
(specified by a list of candidates and a list of voters' preferences
over the candidates).  Bartholdi, Tovey, and
Trick~\cite{bar-tov-tri:j:who-won} proved $\dw$ to be NP-hard.
They posed as an open question the issue of whether $\dw$ is
NP-complete.  
Hemaspaandra, Hemaspaandra, and Rothe~\cite{hem-hem-rot:cbutwithTRUR:dodgson}
recently raised the lower bound of $\dw$
to $\parallelnp$-hardness, thus pinpointing to the exact computational
complexity of checking the winner in Carroll elections (as this
lower bound matches the trivial $\parallelnp$ upper bound).  This also
shows that $\dw$ cannot be NP-complete unless the polynomial hierarchy
collapses, answering the above question raised
by Bartholdi, Tovey, and Trick~\cite{bar-tov-tri:j:who-won}.
The claim that this is a {\em natural\/} $\parallelnp$-complete
problem is 
compelling here as elections are exactly the type of setting in 
which comparisons (seeing who did best) are completely 
natural.  It is also nice that the problem predates by a 
century the class for which it is complete.
Section~\ref{s:carroll} further discusses the complexity 
of Carroll's election system.

Another problem whose lower bound has been recently raised to
$\parallelnp$ is 
{\tt
Minimum} {\tt Equivalent} {\tt Expression} ($\mee$, for short), 
which 
asks:  Given a boolean formula $\phi$ and a positive integer~$k$,
is there a boolean expression having 
at most $k$ occurrences of literals that is equivalent to
$\phi$.  This is
the best-known
form of the problem,
as this version 
is presented and discussed in detail in the widely read 
book of Garey and Johnson~\cite{gar-joh:b:int}.  However,
the problem has a long history in various 
related and contrasting versions,
and is of immense historical importance in complexity 
theory.
In their seminal paper defining the
polynomial hierarchy, Meyer and 
Stockmeyer~\cite{mey-sto:c:reg-exp-needs-exp-space}
state explicitly that this is the issue that led to 
their creation of the polynomial hierarchy:
\begin{quote}
``We were first led to extend the class NP by considering 
the language...~which denotes the set of 
well-formed Boolean expressions for which there is no
shorter equivalent expression.''
\end{quote}
Furthermore, in his journal paper on the polynomial hierarchy,
Stockmeyer~\cite{sto:j:poly} uses as his motivating problem the 
language that is the set of pairs $(\phi,k)$ such that $\phi$ 
is a DNF formula and there is a DNF formula $\phi'$ that is 
equivalent to $\phi$ and has at most $k$ occurrences of 
literals.

Returning from our historical digression, we now focus again
on $\mee$ in the form in which it is stated by 
Garey and Johnson.
Note that $\mee$ has a trivial $\np^{\np}$ upper bound,
yet the best previously known lower bound for $\mee$ was only
coNP-hardness.\footnote{ \protect\singlespacing 
Garey and Johnson write ``NP-hardness,'' but by this they 
mean merely NP-{\em{}Turing}-hardness.} 
Hemaspaandra and Wechsung~\cite{hem-wec:t:mee} 
have within the last year raised this lower bound to
$\parallelnp$-hardness.
Due to space limitations, in this survey we will not further 
discuss this problem, but instead we point the 
reader to~\cite{hem-wec:t:mee}.

Section~\ref{s:greedy} 
is about the Minimum Degree Greedy Algorithm (MDG, for short),
which is a well-known
heuristic algorithm for seeking
a large independent set in a graph. 
When does MDG in fact output a maximum independent set---or
at least when is MDG's output
{\em approximately\/} the size 
(i.e., within a certain fixed constant factor)
of a maximum independent set of
the given graph?  One line of research
tries to identify those graph classes for which MDG has a good
approximation ratio (e.g., graphs with
bounded degree or bounded average
degree~\cite{hal-rad:c:greed-is-good}) or for which MDG actually
outputs a maximum independent set (e.g., trees, split graphs,
``well-covered'' graphs, complete $k$-partite graphs,
and complements of $k$-trees, 
see~\cite{bod-thi-yam:j:greedy-for-maximum-independent-sets,tan-tar:j:well-covered-graphs}).
On the other hand, Bodlaender, Thilikos, and
Yamazaki~\cite{bod-thi-yam:j:greedy-for-maximum-independent-sets}
prove that the problem of recognizing (for any fixed rational~$r \geq
1$) whether, for a given graph~$G$, MDG on input $G$ outputs a set of
size at least $1/r$ times the size of a maximum
independent set of~$G$ is a coNP-hard 
problem.  They also provide a $\parallelnp$
upper bound for this recognition problem and leave open the question
of whether their coNP-hardness lower bound can be improved to match
the upper bound (they actually state their 
upper bound as a $\p^{\np}$ upper bound, but it 
obviously is even a $\parallelnp$ upper bound).  Hemaspaandra and
Rothe~\cite{hem-rot:t:max-independent-set-by-greedy} settle this
question by raising this problem's lower bound to
$\parallelnp$-hardness, thus establishing that 
the problem is complete for
$\parallelnp$.

We have 
mentioned the raising to $\parallelnp$-hardness
of a number of known NP-hardness/coNP-hardness 
lower bounds of some quite
natural problems.
In Section~\ref{s:final-remarks}, we discuss the question of what raising
lower bounds from NP-hardness to hardness for parallel access to NP
actually tells us about the computational complexity of a given
problem.  That is, does $\parallelnp$-hardness give any more insight
into the inherent hardness of a problem than an NP-hardness lower
bound?  We will make the case that the answer to this question
depends on which computational model one views
as most natural.
We discuss this issue
for the following computational paradigms: 
deterministic algorithms, 
probabilistic algorithms,
small circuits,
exact counting, unambiguous computation,
and approximate computation.

\section{Determining the Winner in Lewis Carroll's Election System}
\label{s:carroll}

Let us say that an election is 
specified by a list of candidates and a list of
voters' preferences, where each voter must 
have strict preferences over the candidates.
A candidate $c$ is a {\em Condorcet winner\/} of a given 
election
if $c$ defeats (``defeats'' here means: is preferred
by strictly more than half of the voters) each other candidate in
pairwise majority-rule elections. 
This notion dates to research done in the 1700s by
the Marquis de Condorcet
(\cite{con:b:condorcet-paradox}, 
see~\cite{bla:b:polsci:committees-elections}).
Not all elections have Condorcet winners.
The famous Condorcet Paradox
notes that 
when there are more
than two candidates, pairwise majority-rule elections may yield strict
cycles in the aggregate preference even if each voter has non-cyclic
preferences.

As an example, suppose
that three candidates, Clinton ($C$), Dole ($D$), and Perot ($P$), run
  for president, and there are exactly three voters 
and they have the
  following preferences: $\pair{D<C<P}$ (voter~1), $\pair{C<P<D}$
  (voter~2), and $\pair{P<D<C}$ (voter~3). Note that in a pairwise 
majority-rule election $C$ would beat $D$, $P$ would beat $C$, and
$D$ would beat $P$---a strict cycle.  That is, though each of the 
voters is individually rational (i.e., has transitive preferences),
the voters' aggregated preference (``society's preference'')
is irrational!

Lewis
Carroll developed an election system that respects the notion 
of a Condorcet winner, yet that will never create an 
irrational aggregate preference.
Carroll's voting
system~(\cite{dod:unpubMAYBE:dodgson-voting-system}, see
also~\cite{bar-tov-tri:j:who-won,hem-hem-rot:cbutwithTRUR:dodgson})
works as follows.  Each candidate is assigned a score (that we will
call his or her {\em Carroll score\/}), namely, the minimum number of
sequential {exchanges of two adjacent candidates in the voters'
  preference orders} 
needed to make
the given candidate a Condorcet winner.  In particular, any Condorcet
winner (and if one exists, he or she 
is unique) has a Carroll score of~0. 

As an
example, consider the following electorate: $\pair{D<P<C}$ (voter~1),
$\pair{D<C<P}$ (voter~2), $\pair{C<D<P}$ (voter~3), and $\pair{C<P<D}$
(voter~4). Obviously, $P$'s Carroll score is~0, since $P$ is the
Condorcet winner. $C$'s Carroll
score equals~3.  It is at most~3, as $C$ can be made a
Condorcet winner by, e.g., exchanging $C$ upwards in
the preferences of voter~2 (once)
and voter~4 (twice), which yields the new preferences:
$\pair{D<P<C}$, $\pair{D<P<C}$, $\pair{C<D<P}$, and
$\pair{P<D<C}$. 
Also, it is not hard to see
that no two exchanges will make $C$ a Condorcet winner.
(By symmetry, 
$D$ also has a Carroll score of~3.)

A candidate $c$ {\em ties-or-defeats\/} a candidate $d$ if the score
of $d$ is not less than that of~$c$. A candidate $c$ is said to win
a Carroll election if $c$ ties-or-defeats all other candidates.
Of course, due to ties it is possible for two candidates (such as $C$
and $D$ in the above example) to tie-or-defeat each other.  
Thus, some 
elections may have more than one winner. However, it
is important to note that
in Carroll elections no strict-preference cycles are possible, since
each candidate is assigned an integer score.

Bartholdi, Tovey, and Trick~\cite{bar-tov-tri:j:who-won} provided a
lower bound---NP-hardness---on the computational complexity of
determining the election winner(s) in Carroll's system, a problem we
will call $\dw$. Formally, $\dw$ is the problem of determining, given a
set $C$ of candidates, a distinguished member $c$ of $C$, and a
list $V$ of the voters'
preference orders on $C$, whether $c$ ties-or-defeats
all other candidates in the election.  
Bartholdi, Tovey, and Trick leave open the problem of
what the exact computational complexity of $\dw$ is.
Hemaspaandra,
Hemaspaandra, and Rothe~\cite{hem-hem-rot:cbutwithTRUR:dodgson}
solve this question by proving $\dw$ to be
$\parallelnp$-complete. 

The upper bound can easily be seen as follows. First note that for a
given candidate $c$ and a given integer~$k$, it can be decided in NP
whether $c$'s score is at most $k$~\cite{bar-tov-tri:j:who-won}.
Also, given an election as input, the
number of candidates and the highest possible score for each candidate
are both polynomially bounded in the input length. Hence, we can in
parallel ask to an NP set all plausible Carroll scores for each of the
given candidates and thus can compute the exact Carroll score for each
candidate. After having done so, it is easy to decide whether or not
the designated candidate $c$ ties-or-defeats all other candidates in
the election.  This easy argument establishes the upper bound.

Unfortunately,
the proof of the lower bound is too long and complicated to be
included here (interested readers are referred
to~\cite{hem-hem-rot:cbutwithTRUR:dodgson}). A short,
handwaving outline of the argument's flavor
follows.  

The general proof strategy is as follows:  In order to 
prove that $\dw$ is {\em hard\/} ($\parallelnp$-hard),
Hemaspaandra,
Hemaspaandra, and Rothe build a broad set of polynomial-time
algorithms for manipulating (in the common English-language 
sense, not the term-of-art political science sense) 
Carroll elections.  In particular, they build 
algorithms that manipulate the parity of elections,
that allow groups of elections to be ``summed'' 
in such a way that the score of the sum equals the sum of the 
scores, and that allow elections to be merged in such a way as to 
preserve information yet avoid interference.
Using these algorithms, Wagner's toolkit, and going through some 
intermediate problems and arguments, the $\parallelnp$ lower
bound is established.

\section{Greed is Hard to Understand: Approximation
and Maximum Independent Sets}\label{s:greedy}

The Minimum Degree Greedy Algorithm (MDG) chooses some vertex of
minimum degree from the input graph, adds this vertex to its output
set, deletes all neighbors of this vertex, and repeats this procedure
with the accordingly updated graph until an empty graph is left. How
well does MDG approximate a maximum independent set of the given graph
(i.e., a maximum-sized 
subset $S$ of the vertices of the graph such that for
no pair of vertices in $S$ is there an edge connecting them)? We
denote the size of the maximum independent sets of graph $G$
by~$\alpha(G)$. Let $\mbox{mdg\/}(G)$ denote the maximum size of the
output set of MDG on input~$G$, where the maximum is taken over all
the possible 
choices MDG has among the vertices
of minimum degree when there is more than one such vertex.

Bodlaender, Thilikos, and
Yamazaki~\cite{bod-thi-yam:j:greedy-for-maximum-independent-sets}
define (for any fixed rational~$r \geq 1$) the class of graphs for
which MDG, taking a best possible sequence of choices, approximates
the size of a maximum independent set within a constant factor
of~$r$, i.e., for which $\alpha(G)/r \leq \mbox{mdg}(G)$. They
denote this recognition problem by ${\cal S}_r$ and prove that for any
rational~$r \geq 1$, ${\cal S}_r$ is coNP-hard and is contained in
$\parallelnp$. 
They leave open the question of whether the lower and/or
the upper bound can be improved. For the special case of $r = 1$, they
show that ${\cal S}_1$ even is DP-hard, where DP~\cite{pap-yan:j:dp}
denotes the class of sets that can be represented as the difference of
two NP sets (DP clearly contains both NP and coNP and is 
equal to neither  NP nor coNP unless the 
polynomial hierarchy
collapses~\cite{kad:joutdatedbychangkadin:bh}), 
again leaving open the issue of
whether this lower bound can be raised.

Hemaspaandra and 
Rothe~\cite{hem-rot:t:max-independent-set-by-greedy} settle 
these questions by establishing the exact computational
complexity of recognizing all these graph classes: For each rational
$r \geq 1$, ${\cal S}_r$ is $\parallelnp$-complete.   The upper
bound is implicit in the proof 
of~\cite[Lemma~6]{bod-thi-yam:j:greedy-for-maximum-independent-sets}.
Namely, to determine whether a given graph with $n$
vertices does {\em not\/} belong to ${\cal S}_r$, it is sufficient to
ask whether for some~$k$, $1 \leq k \leq n$, it holds that
(i)~$\alpha(G) \geq k$, and (ii)~for no possible sequence of choices does
MDG on input $G$ output a set of size at least $k/r$, i.e.,
$\mbox{mdg}(G) < k/r$. Note that all these queries 
are NP-type or coNP-type queries, and can be 
resolved via parallel access to the NP-complete set SAT
(using the standard complementation-of-the-answer 
trick for the coNP-type questions).
This algorithm places the complement of ${\cal
S}_r$, and thus ${\cal S}_r$ itself, into~$\parallelnp$.

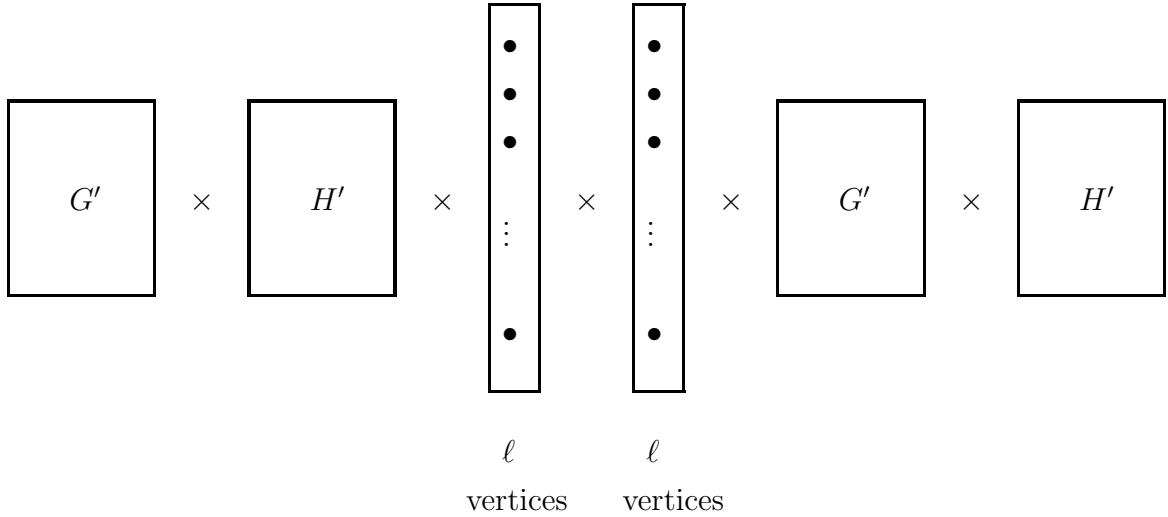
\begin{figure}
\begin{center}
\setlength{\unitlength}{0.00083300in}%
\begingroup\makeatletter\ifx\SetFigFont\undefined%
\gdef\SetFigFont#1#2#3#4#5{%
  \reset@font\fontsize{#1}{#2pt}%
  \fontfamily{#3}\fontseries{#4}\fontshape{#5}%
  \selectfont}%
\fi\endgroup%
\begin{picture}(7244,3199)(1179,-2938)
\thicklines
\put(2701,-1561){\framebox(900,1200){}}
\put(4201,-2161){\framebox(300,2400){}}
\put(5101,-2161){\framebox(300,2400){}}
\put(6001,-1561){\framebox(900,1200){}}
\put(7501,-1561){\framebox(900,1200){}}
\put(1576,-1036){\makebox(0,0)[lb]{\smash{\SetFigFont{12}{14.4}{\rmdefault}{\mddefault}{\updefault}$G'$}}}
\put(3076,-1036){\makebox(0,0)[lb]{\smash{\SetFigFont{12}{14.4}{\rmdefault}{\mddefault}{\updefault}$H'$}}}
\put(6376,-1036){\makebox(0,0)[lb]{\smash{\SetFigFont{12}{14.4}{\rmdefault}{\mddefault}{\updefault}$G'$}}}
\put(7876,-1036){\makebox(0,0)[lb]{\smash{\SetFigFont{12}{14.4}{\rmdefault}{\mddefault}{\updefault}$H'$}}}
\put(2326,-1036){\makebox(0,0)[lb]{\smash{\SetFigFont{12}{14.4}{\rmdefault}{\mddefault}{\updefault}$\times$}}}
\put(3826,-1036){\makebox(0,0)[lb]{\smash{\SetFigFont{12}{14.4}{\rmdefault}{\mddefault}{\updefault}$\times$}}}
\put(4726,-1036){\makebox(0,0)[lb]{\smash{\SetFigFont{12}{14.4}{\rmdefault}{\mddefault}{\updefault}$\times$}}}
\put(5626,-1036){\makebox(0,0)[lb]{\smash{\SetFigFont{12}{14.4}{\rmdefault}{\mddefault}{\updefault}$\times$}}}
\put(1201,-1561){\framebox(900,1200){}}
\put(7126,-1036){\makebox(0,0)[lb]{\smash{\SetFigFont{12}{14.4}{\rmdefault}{\mddefault}{\updefault}$\times$}}}
\put(4276,-1861){\makebox(0,0)[lb]{\smash{\SetFigFont{12}{14.4}{\rmdefault}{\mddefault}{\updefault}$\bullet$}}}
\put(4276,-2611){\makebox(0,0)[lb]{\smash{\SetFigFont{12}{14.4}{\rmdefault}{\mddefault}{\updefault}$\ell$}}}
\put(5176,-2611){\makebox(0,0)[lb]{\smash{\SetFigFont{12}{14.4}{\rmdefault}{\mddefault}{\updefault}$\ell$}}}
\put(5026,-2911){\makebox(0,0)[lb]{\smash{\SetFigFont{12}{14.4}{\rmdefault}{\mddefault}{\updefault}vertices}}}
\put(4051,-2911){\makebox(0,0)[lb]{\smash{\SetFigFont{12}{14.4}{\rmdefault}{\mddefault}{\updefault}vertices}}}
\put(4276,-61){\makebox(0,0)[lb]{\smash{\SetFigFont{12}{14.4}{\rmdefault}{\mddefault}{\updefault}$\bullet$}}}
\put(4276,-361){\makebox(0,0)[lb]{\smash{\SetFigFont{12}{14.4}{\rmdefault}{\mddefault}{\updefault}$\bullet$}}}
\put(4276,-661){\makebox(0,0)[lb]{\smash{\SetFigFont{12}{14.4}{\rmdefault}{\mddefault}{\updefault}$\bullet$}}}
\put(5176,-1861){\makebox(0,0)[lb]{\smash{\SetFigFont{12}{14.4}{\rmdefault}{\mddefault}{\updefault}$\bullet$}}}
\put(5176,-61){\makebox(0,0)[lb]{\smash{\SetFigFont{12}{14.4}{\rmdefault}{\mddefault}{\updefault}$\bullet$}}}
\put(5176,-361){\makebox(0,0)[lb]{\smash{\SetFigFont{12}{14.4}{\rmdefault}{\mddefault}{\updefault}$\bullet$}}}
\put(5176,-661){\makebox(0,0)[lb]{\smash{\SetFigFont{12}{14.4}{\rmdefault}{\mddefault}{\updefault}$\bullet$}}}
\put(5176,-1261){\makebox(0,0)[lb]{\smash{\SetFigFont{12}{14.4}{\rmdefault}{\mddefault}{\updefault}$\vdots$}}}
\put(4276,-1261){\makebox(0,0)[lb]{\smash{\SetFigFont{12}{14.4}{\rmdefault}{\mddefault}{\updefault}$\vdots$}}}
\end{picture}
\end{center}
\caption{\label{fig:reduction} Reducing
  {\tt MIS}$_{=}$ to ${\cal S}_1$: Graph~$\widehat{G}$ constructed
  from the graphs $G'$ and~$H'$.}
\end{figure}

For the lower bound, we will here describe only one special case:  the
proof that ${\cal S}_1$ in fact is $\parallelnp$-hard, which improves
upon the previously known DP lower bound. This special case of 
the reduction is
short enough to fit into a 
footnote.\footnote{ \protect\singlespacing We reduce the problem {\tt
    MIS}$_{=}$ to ${\cal S}_1$, where {\tt MIS}$_{=}$ is defined to be
  the set of all pairs $(G,H)$ of graphs $G$ and $H$ such that
  $\alpha(G) = \alpha(H)$. By Wagner's work~\cite{wag:j:more-on-bh}
  (see also~\cite{hem-rot:t:max-independent-set-by-greedy}),
   {\tt
    MIS}$_{=}$ is $\parallelnp$-complete.
So let a pair of graphs, $(G,H)$, be given.
It is easy to see that without loss of generality we can assume that $G$ and $H$
have the same number, $k$, of edges.
Now we use the
construction
of~\cite[Theorem~4]{bod-thi-yam:j:greedy-for-maximum-independent-sets}
to transform $G$ and $H$ into two new graphs $G'$ and
$H'$ such that: (i)~$G'$ and $H'$ both are in
${\cal S}_1$, (ii)~$\alpha(G') = \alpha(G) + k$, and
(iii)~$\alpha(H') = \alpha(H) + k$.  Let $\ell$ be some integer larger than
the larger 
of the number of vertices in $G'$ and the number of vertices in $H'$.
The actual construction is as follows. 
Take two copies of $G'$, two copies
of $H'$, and two copies
of a set consisting of $\ell$ isolated new vertices, and connect
these six subgraphs 
as shown in Figure~\ref{fig:reduction}, where a ``$\times$''
between two subgraphs denotes their Cartesian product, i.e., any two
vertices $u$ and $v$ (where $u$ is from the one subgraph and $v$ is 
from the other)
are joined by an edge. Call the resulting graph~$\widehat{G}$.
Since $G' \in {\cal S}_1$ and $H' \in {\cal S}_1$, we have
$\mbox{mdg}(\widehat{G}) = \alpha(G') + \alpha(H') + \ell 
                        = \alpha(G) + \alpha(H) + 2k + \ell.$
On the other hand,
$\alpha(\widehat{G}) = 2 \cdot \max\{\alpha(G') , \alpha(H')\}
                         + \ell 
                  = 2 \cdot \max\{\alpha(G) , \alpha(H)\} + 2k + \ell.$
Thus, $\widehat{G} \in {\cal S}_1$ if and only if 
$\alpha(G) = \alpha(H)$.}
Interested readers are referred
to~\cite{hem-rot:t:max-independent-set-by-greedy} for the more
complicated proof of the general case, i.e., 
the proof for arbitrary rationals $r \geq 1$.

\section{Final Remarks: Lower Bounds and 
Differing Forms of Computation}\label{s:final-remarks}

In this final section, we will be dealing with the question: What
impact does raising some problem $FOO$'s 
lower bound from NP-hardness to
$\parallelnp$-hardness actually have regarding the problem's
potential solvability via other models of computation (as captured 
by their respective complexity classes).
By the definition of hardness, this can be formally
phrased as follows: {\em If ${\cal C}$ is some complexity class, is
  it currently known to hold that $\np \subseteq {\cal C}$ if and only if
  $\parallelnp \subseteq {\cal C}$\/}? An affirmative answer 
means that the raised lower bound is worthless with regard
to~${\cal C}$.  A negative answer, however, implies that the raised
lower bound may have some value in the model captured by~${\cal C}$
(on the other hand, it might just be the case that 
$\np \subseteq {\cal C} \iff
\parallelnp \subseteq {\cal C}$ truly holds and
researchers have simply to date failed  to establish that it holds).
That is, if the implication $\np \subseteq {\cal C} \implies
\parallelnp \subseteq {\cal C}$ is not 
currently known to hold, then in light of the
problem $FOO$'s new $\parallelnp$ lower bound, placing $FOO$ into
${\cal C}$ gives us a potentially 
stronger conclusion than what was previously
known: $\parallelnp \subseteq {\cal C}$ instead of merely $\np
\subseteq {\cal C}$.  In other words, we may 
take ``$FOO$ is $\parallelnp$-hard'' to be potentially stronger
evidence that $FOO$ cannot be solved by the computational 
power modeled by $\cal C$ (i.e., that $FOO\not\in {\cal C}$)
than was available from the fact that ``$FOO$ is NP-hard.''
(The converse implication, $\parallelnp
\subseteq {\cal C} \implies \np \subseteq {\cal C}$, is trivial
and so needs no discussion.)

We mention, however, that independent of the 
``connections to other computational models'' issues discussed 
in this section,
raising from NP-hardness to $\parallelnp$-hardness
the lower bounds of such natural and
longstanding problems as those we have discussed
is a genuine improvement in terms of placement
within the polynomial hierarchy (unless the polynomial
hierarchy itself collapses).

In light of the above discussion, we now discuss for 
various computational models (as captured by their complexity
classes ${\cal C}$) whether: $\np \subseteq {\cal C} 
\iff \parallelnp \subseteq {\cal C}$.

\smallskip

{\bf Deterministic Polynomial Time (P).} \quad The class P is so low
in power that from its viewpoint, it does not matter at
all whether a certain problem has a $\parallelnp$ lower 
bound or just an NP
lower bound, since clearly $\np = \p$ if and only if $\parallelnp =
\p$.  Of course, this fact has been well-known since the 
seminal paper of 
Meyer and Stockmeyer~\cite{mey-sto:c:reg-exp-needs-exp-space},
where $\p = \np \iff \p = \ph$ is explicitly noted.

\smallskip

{\bf Probabilistic Polynomial Time with unbounded and bounded
  two-sided error (PP and BPP).} \quad 
PP~\cite{sim:thesis:complexity,gil:j:probabilistic-tms}
(respectively,
BPP~\cite{gil:j:probabilistic-tms}) is defined to be the class of
languages $L$ for which there exists a probabilistic polynomial-time
Turing machine $M$ such that, for all inputs~$x$, if $x \in L$ then $M$
accepts its input $x$ with probability~$\geq 1/2$ 
(respectively,~$\geq 3/4$), and if $x \not\in L$ then $M$ accepts its input $x$
with probability~$< 1/2$ (respectively,~$\leq 1/4$).  Clearly, $\bpp
\subseteq \pp$ and $\np \cup \conp \subseteq \parallelnp$.
It is known that $\parallelnp
\subseteq
\pp$~\cite{bei-hem-wec:j:powerprob}.
This latter inclusion immediately implies that $\np \subseteq \pp$ if and only
if $\parallelnp \subseteq \pp$, since both are 
outright true.  BPP and NP probably are incomparable.  However,
since BPP is closed under Turing reductions, 
we easily have: $\np \subseteq \bpp$
if and only if $\parallelnp \subseteq \bpp$.\footnote{
  \protect\singlespacing It is even known that $\np \subseteq \bpp$ if
  and only if the entire polynomial hierarchy
  is
  contained in BPP~\cite{hel-zac:j:decisive}.}  
Thus, raising a set's lower bound from NP-hardness to $\parallelnp$-hardness
does not in and of itself give one any heightened level of evidence that
the problem is not in BPP or is not in PP.

\smallskip

{\bf Randomized Polynomial Time with one-sided and zero-sided error (R
  and ZPP).}  \quad 
R~\cite{gil:j:probabilistic-tms}
is defined to be the class of languages $L$ for which there exists a
probabilistic polynomial-time Turing machine $M$ such that, for all
inputs~$x$, if $x \in L$ then $M$ accepts its input $x$ with
probability at least $1/2$, and if $x \not\in L$ then $M$ accepts its
input $x$ with probability~0. Clearly, $\p \subseteq \rp \subseteq
\np$ and $\rp \subseteq \bpp$. Like~NP, the class R is not known to be
closed  
under Turing reductions or even under complementation---R
and coR perhaps differ. $\zpp$~\cite{gil:j:probabilistic-tms}
equals the class of languages that can be solved in 
expected polynomial time, and it is known
that $\zpp = \rp \cap \corp$.
Since ZPP (like BPP) is closed 
under Turing reductions, the above comment about BPP 
applies analogously 
to~ZPP:  $\np = \zpp$ if and only if $\parallelnp = \zpp$.
Thus, raising a set's lower bound from NP-hardness to $\parallelnp$-hardness
does not in and of itself give one any heightened level of evidence that
the problem is not in ZPP.

In contrast, it is not known whether $\np = \rp$ implies $\parallelnp
= \rp$.  (The best result known in this direction is that $\np = \rp$
implies $\parallelnp \subseteq \bpp$, due to the fact that BPP is
closed  under Turing reductions and $\rp \subseteq \bpp$.)
Thus, raising a set's lower bound from NP-hardness to $\parallelnp$-hardness
potentially provides a heightened level of evidence that
the problem is not in R\@.  
(Of course, if one believes as an article
of faith that $\rp \neq \np$ then this heightening claim is not 
applicable, as in that case 
one cannot believe that even one NP-hard set might be in R.)

\smallskip

{\bf Exact Counting ({\boldmath $\ceqp$}).} \quad
$\ceqp$~\cite{sim:thesis:complexity,wag:j:succinct} is the class of
sets $L$ such that there is a polynomial-time function $f$
and a nondeterministic polynomial-time Turing machine $N$ such that,
for each $x$, $x\in L$ if and only if $N$ on input $x$ has exactly
$f(x)$ accepting paths. It is well-known that $\conp \subseteq \ceqp
\subseteq \pp$~\cite{sim:thesis:complexity,wag:j:succinct}.  
Like~R, $\ceqp$ is neither known to be closed under
Turing reductions nor known to be closed under complementation.
We claim that 
$\np \subseteq \ceqp \implies \parallelnp \subseteq \ceqp$
nonetheless holds.  We will prove this using
other closure properties that $\ceqp$ is known to possess.
\begin{theorem}\label{t:thma}
$\np \subseteq \ceqp \iff \parallelnp \subseteq \ceqp$.
\end{theorem}

\noindent
{\bf Proof: \quad}
Assuming $\np \subseteq \ceqp$
  and recalling $\conp \subseteq \ceqp$, we have that 
DP (see Section~\ref{s:greedy})
is contained in $\ceqp$, since each DP set is the 
intersection of an NP set and a coNP set, and 
Gundermann, Nasser, and Wechsung~\cite{gun-nas-wec:c:counting-survey} 
have shown that 
$\ceqp$ is closed under intersection. 
Since $\ceqp$ is 
known (\cite{gun-nas-wec:c:counting-survey}, see 
the discussion in~\cite{rot:technicalreport:some-closure}
and~\cite{bei-cha-ogi:j:difference-hierarchies})
to also be 
 closed
  under disjunctive truth-table reductions~\cite{lad-lyn-sel:j:com}
 and since the disjunctive truth-table
  closure of DP is equal to
  $\parallelnp$
it follows that
  $\parallelnp \subseteq \ceqp$.
Regarding the claim we just made that 
$ \{ L \condition  (\exists A \in {\rm DP}) [L \leq_{dtt}^p A]\}
= \parallelnp$,
or equivalently, using ``R'' notation,
$$ \red{{dtt}}{{p}}{{{\rm DP}}} = \parallelnp,$$
we claim that this is implicit and immediate from the
fact that a certain
set known as ${\rm PARITY}_{\omega}^{\rm SAT}$
that Buss and Hay~\cite{bus-hay:j:tt}
proved $\parallelnp$-complete 
is clearly
in
$\red{{dtt}}{{p}}{{{\rm DP}}}$.~\qed

\smallskip

{\bf Unambiguous Polynomial Time (UP).} \quad UP~\cite{val:j:checking}
is the class of those NP sets that are accepted 
via some NP machine that, on
each input, has at most one accepting path. Clearly, $\p \subseteq \up
\subseteq \np$. Like R and $\ceqp$, UP is not known to be closed under
Turing reductions (or even under complementation).  Unlike $\ceqp$,
however, UP (though clearly closed under intersection) is not known to
possess any other useful closure properties that might be exploited
instead. Thus, it is an open question
whether $\np = \up$ implies $\parallelnp =
\up$.  This leaves open the possibility that raising NP lower bounds to
$\parallelnp$-hardness is an improvement in terms of giving 
evidence that a given problem is not 
in UP\@.  
(Of course, if one believes as an article
of faith that $\up \neq \np$ then this heightening claim is not 
applicable, as in that case 
one cannot believe that even one NP-hard set might be in UP.)

\smallskip

{\bf Small circuits (P/poly) and Approximation Models (P-close and
  APT).} \quad P/poly denotes the class of all sets that can be
decided by polynomial-size circuits. By a result of Meyer
(reported 
in~\cite{ber-har:j:iso}), a set $A$ is in P/poly if and only if $A \in
\p^{S}$ for some sparse set~$S$.\footnote{ \protect\singlespacing A
  set $S$ is said to be {\em sparse\/} if there is a polynomial $p$
  such that, for all lengths $n$, the number of elements of $S$ up to
  length~$n$ is bounded by~$p(n)$.}  Thus, P/poly is clearly closed
under Turing reductions, which gives: 
$\np \subseteq \ppoly$ if and only if $\parallelnp \subseteq
\ppoly$.

A set $A$ is {$\p$-close\/}~\cite{sch:j:closeness} 
if there is a P set $B$ such
that the symmetric difference of $A$ and $B$ is a sparse set.  That
is, each P-close set in a sense is ``approximated''
by a P set.
However, Sch\"oning~\cite{sch:j:closeness} 
proved that if every NP set is P-close, then $\np =
\p$, 
which in turn implies that every $\parallelnp$ set is
P-close. Thus, we have:  $\np 
\subseteq {\rm P\mbox{-}close}$ if and only if 
$\parallelnp 
\subseteq {\rm P\mbox{-}close}$.
An interesting special case of P-closeness
is provided by the class APT (almost polynomial
time)~\cite{mey-pat:t:int}. APT is the class of sets having
deterministic algorithms that run in polynomial time for all inputs
except those in a sparse set. Since every APT set is P-close, the
above result of Sch\"oning easily gives:
$\np 
\subseteq {\rm APT}$ if and only if 
$\parallelnp 
\subseteq {\rm APT}$.

Stepping back to summarize the contents of this section and 
this article:
A number of recent results improve from NP-hardness
(or coNP-hardness) to
$\parallelnp$-hardness the lower bounds for 
natural problems whose computational complexities
have long been open issues.    
In fact, two of these natural problems are complete
for $\parallelnp$, lending credibility to the naturalness of 
the class $\parallelnp$.  However, as a cautionary note, 
we pointed out that many other
computational modes are so sharply orthogonal to complexity 
as measured in the polynomial hierarchy that raising lower bounds in
the polynomial hierarchy does not speak directly to raising 
complexity in these other computational modes.

\smallskip
{\bf Acknowledgments:}~~We thank Gerd Wechsung for pointing out
the DP claim in the proof of Theorem~\ref{t:thma}.

\bibliography{gry}

\end{document}